\documentclass[preprint]{aastex}

\slugcomment{\it Submitted to The Astrophysical Journal}

\begin{document}

\title{UV Imaging of the Galaxy Cluster CL0939$+$4713 (Abell~851)
    at z=0.41\altaffilmark{1}}

\altaffiltext{1}{Based on observations with the NASA/ESA Hubble Space
    Telescope, obtained at the Space Telescope Science Institute, which is
    operated by AURA, Inc., under NASA Contract NAS 5-26555.}

\author{Lucio M. Buson\altaffilmark{2}, Francesco Bertola\altaffilmark{3},
Michele Cappellari\altaffilmark{3}, Cesare Chiosi\altaffilmark{3},
Alan Dressler\altaffilmark{4}, and Augustus Oemler, Jr.\altaffilmark{4}}

\altaffiltext{2}{Osservatorio di Capodimonte, Napoli, Italy}
\altaffiltext{3}{Dipartimento di Astronomia, Universit\`a di Padova, Italy}
\altaffiltext{4}{Carnegie Observatories, USA}

\begin{abstract}

The first UV F300W and F218W WFPC2 observations of the rich galaxy
cluster CL0939$+$4713 at z~=~0.41 are presented and discussed.
UV/optical two-color and c-m diagrams of the sources detected in the
F300W waveband are constructed. Thanks to pre-existing HST optical
images of the same field a morphological classification for the majority
of these objects is also provided. Moreover, taking advantage of recent
redshift surveys along CL0939$+$4713 line of sight, separate diagrams
comparing the properties of galaxies belonging to the cluster and to its
close projected field are presented. Possible evolutionary effects in
the UV from z $\sim$0.4 to present time are investigated by comparing
the restframe [mid-UV$-$Optical] colors of galaxies in CL0939$+$4713
with balloon-borne data of the Coma cluster, as well as by resorting to
suitable galaxy evolution models. Finally, current attempts to constrain
the epoch of the UV-upturn onset in evolved populations by means of HST
UV observations are discussed.

\end{abstract}


\clearpage

\section{INTRODUCTION}

Owing to its high richness and distance (z=0.41), CL0939$+$4713 =
Abell~851 turns out to be one of the ``most wanted'' targets both for
morphological and evolutionary studies of galaxy clusters, thus being
presently one of the best known intermediate-redshift systems. Extensive
analyses, including both imaging and spectroscopy, have been carried out
from ground in the optical (Dressler \& Gunn 1983; 1992; Fukugita {\em
et al.} 1995; Belloni {\em et al.} 1995; Belloni \& R\"oser 1996;
Dressler {\em et al.} 1999) and in the IR (Stanford {\em et al.} 1995,
1998). Besides the above ground-based studies, a large amount of
space-borne (HST, ROSAT) observing time has been devoted to survey this
cluster (Dressler {\em et al.} 1994a, 1994b, 1997; Seitz {\em et al.}
1996; Oemler {\em et al.} 1997; Smail {\em et al.} 1997; Andreon {\em et
al.} 1997; Schindler \& Wambsganss 1996; Schindler {\em et al.} 1998).

An important outcome of such -- mainly optical -- observational effort
is the identification in CL0939$+$4713 (as well as in the cores of other
rich clusters of comparable redshift) of a larger fraction of
star-forming galaxies. Such a phenomenon, commonly referred to as the
Butcher--Oemler effect (Butcher \& Oemler 1978; 1984), is related, in
turn, to an excess of late--type spirals, irregulars and mergers, when
compared with the nearby cluster galaxy population (e.g. Couch {\em et
al.} 1994, 1998; Oemler {\em et al.} 1997). Analogously, a remarkably
higher proportion of the post-star-forming a$+$k and/or k$+$a galaxies
(previously termed a bit improperly E$+$A galaxies) can be recognized in
these intermediate redshift clusters (Dressler {\em et al.} 1999;
Poggianti {\em et al.} 1999). However, also notable is the presence of a
well developed population of `old' stellar systems as a substantial
fraction of the luminous ellipticals that dominate clusters today seem
to be well in place by z$\sim$0.4.

The unique HST imaging capabilities at {\em ultraviolet} wavelengths
allow us now to complement the above observations by exploring for the
first time the {\em rest-frame}, far- and mid-UV portion of galaxy SEDs
in this rich cluster. The importance of such data can be appreciated
taking into account that mid-UV colors provide tighter constraints to
the evolution in time and composition of aging stellar populations than
optical-band indices do (cfr. Dorman \& O'Connell 1997). What is more,
the far-UV region (and -- at a lower level -- the mid-UV itself)
represent direct probes of star-formation activity and -- through the
UV-upturn phenomenon -- an additional clue to the composition and age of
old populations. The relative intensity of the far-UV flux changes
rapidly both for young {\em and} old stellar systems indeed, according
either to the fast evolution of recently formed UV-bright stars or the
sudden onset of hot {\em evolved} components like those observed in
present-day stellar populations of elliptical galaxies.

The very limited access to the UV imaging in the past years heavily
hampered this kind of studies, however. The very few papers available in
the literature make use of the datasets provided by the FOCA
balloon-borne Telescope (Donas {\em et al.} 1995, 1997) and the
Ultraviolet Imaging Telescope (UIT) during the Astro-1, Astro-2 missions
onboard Space Shuttle (Cornett {\em et al.} 1998). These analyses could
not be extended to clusters farther than z=0.23, while the approach
recently adopted by Brown {\em et al.} (1998) to get deep HST data of
the ellipticals in the cluster Abell~370 at z=0.375 is in turn hampered
by the small size of the FOC field.

\section{OBSERVATIONS AND REDUCTIONS}

Ultraviolet images of CL0939$+$4713 were obtained with the WFPC2 on January,
14-15, 1996 (Programme ID: 5919). A total of five orbits (2800~s exposure time
each) were spent through the near-UV filter F300W (recording the
cluster's restframe mid-UV with bandwidth $\Delta\lambda\sim$520~\AA\ and
central wavelength $\lambda_c\sim$2100~\AA), while ten additional orbits (again
2800~s exposure time) were devoted to observations through the less sensitive
filter F218W ($\lambda_c\sim$1550~\AA\ and $\Delta\lambda\sim$280~\AA\ at
z=0.41). The adopted WFPC2 field was approximately the same as that chosen for
the pre-existing, optical ($\lambda_c\sim$4870~\AA\ and
$\Delta\lambda\sim$980~\AA\ at cluster distance) F702W images (cf. Dressler
{\em et al.} 1994b). Our UV images and newly in-flight calibrated, dearchived
optical frames were separately aligned, co-added and properly cleaned from
cosmic ray signatures. After separately subtracting a constant sky level to
each WFPC2 chip (this is made possible also for F702W frames by the higher
quality flat-fielding of the data provided by the current pipeline), we
performed aperture photometry on the F300W, F218W and F702W final images.
Standard HST magnitudes (STMAG) $m_{300}$, $m_{702}$ of each obvious source
identified in our UV F300W image have been derived, together with $m_{218}$
magnitudes for the two sources identifiable at shorter wavelengths (the flux,
within the adopted aperture of 0\farcs6 in radius, is at least $5\sigma$ above
the noise, for all chosen sources). Such an aperture gives the lowest scatter
in our derived magnitudes and represents a good compromise between the need of
picking up as much signal as possible and avoiding the danger of including more
than one object. All UV/optical colors used later in our CMDs refer to the
above aperture.

\section{RESULTS}

For each object detected in the near-UV final image F300W (sixty out of
the 181 classified in the optical by Dressler {\em et al.} 1994b) we
provide in Table~1 our measured UV STMAG $m_{300}$, together with the
UV/optical color ($m_{300}-m_{702}$). Analogously, for the two F218W
detected galaxies one can derive a STMAG $m_{218}$, both colors
($m_{218}-m_{300}$) and ($m_{218}-m_{702}$) are provided in Table~2.

Near-UV and optical WFPC2 images are shown in Fig.~1 and Fig.~2, respectively.
The smaller PC1 field has been excluded from our analysis, due to the lower
signal-to-noise it provides. Observed bright objects in the restframe mid-UV
(F300W; $\lambda$$\sim$2100~\AA) are typically spiral galaxies (often with the
majority of the UV flux coming from ultra-luminous sites of star formation
along spiral arms) as well as a few irregular and/or merging systems, while the
optically most prominent giant ellipticals virtually disappear at UV
wavelengths. However, one should be put on his guard against believing that the
totality of spirals identified by HST in the optical can be easily detected in
the mid-UV (F300W) waveband. The majority of such objects escape detection,
instead. The reason why they are not conspicuous in the UV could be related
either to their low star-forming activity (for those belonging to the k and
k$+$a classes of Dressler {\em et al.} 1999), or to the heavy effect of
internal absorption by dust in the UV.

The overall HST c-m diagram ($m_{702}$ {\em vs.} $m_{300}-m_{702}$) of
the sources detected at UV wavelengths is given in Fig.~3. Here the
($m_{300}-m_{702}$) color derived from our aperture photometry (0\farcs6
radius) is compared with {\em integrated} $m_{702}$ magnitudes given by
Smail {\em et al.} (1997). The reason is the latter magnitudes are
representative of the {\em total} optical flux of the galaxies, whereas
our aperture photometry, though well-suited to give reliable
(UV$-$Optical) colors, is not. Specific diagrams within the same figure
isolate galaxies for which -- on the basis of the spectroscopic and
spectrophotometric redshifts provided by Dressler {\em et al.} (1999)
and Belloni and R\"oser (1996) -- one can {\em firmly establish the
association} either with CL0939$+$4713 or the projected field,
respectively. As far as the membership to CL0939$+$4713 is concerned, a
conservative criterion, namely $\Delta$z=0.39--0.42, essentially
matching that of Dressler {\em et al.}, has been adopted.

By cross-identifying our UV sources and the galaxies optically
classified within the same WFPC2 field by Smail {\em et al.} (1997) one
can ascribe a broad morphological classification (namely, E/SO, Spiral
or Irregular) to the majority of objects detected in our F300W image.
This, in turn, allows us to draw some basic conclusions about the
appearance of these UV-bright members of each class. While spirals
clearly span quite a large range in optical luminosity {\em and}
UV/optical color ($\approx$~5~mag in both $m_{702}$ and
[$m_{300}-m_{702}$] color), bright ($m_{702}<$22) early-type galaxies
are always fairly red ([$m_{300}-m_{702}]\sim$3) and detectable
irregular galaxies are constantly bluer than ($m_{300}-m_{702})=$~0. As
far as this behavior is concerned, no significant differences emerge
when comparing confirmed cluster and field galaxies.

A separate issue is represented by fainter, {\em particularly UV-bright}
compact sources falling in the lower left region of the diagram and
whose morphology cannot be reliably identified. These objects presumably
represent quite a heterogeneous zoo, including individual clumps of
vigorous star formation within amorphous, low-luminosity galaxies and
mergers belonging to (or projected on) CL0939$+$4713, as well as
background AGNs. The latter case is true for our bright source no.~1 in
Table~1, identified as a background QSO at z$\sim$2 (Dressler {\em et
al.} 1993; Hutchings \& Davidge 1997).

The availability of ground-based ($g-r$) optical colors for a large fraction of
objects (both cluster members and non-members) detected in our near-UV image
offers also the opportunity of constructing {\em two-color}, optical {\em vs.}
UV/optical (i.e. [$g-r$] {\em vs.} [$m_{300}-m_{702}$]) diagrams (see Fig.~4).
The above ($g-r$) colors come from the ground-based photometry of Dressler \&
Gunn (1992) and are widely discussed also in Dressler {\em et al.} (1994a;
1994b). HST-based and ground-based colors appear well-correlated in Fig.~4, in
such a way that, {\em a posteriori}, one can assume ($g-r$) colors are highly
predictive of the $m_{300}-m_{702}$ colors of galaxies, too. Moreover, no
evidence of a dichotomy between spirals experiencing steady-state
star-formation and possible, UV-dominated spirals undergoing major starbursts
is seen.

\section{COMPARISON WITH A PRESENT-DAY CLUSTER POPULATION}

As stressed above, the restframe mid-UV region -- besides its higher
sensitivity to the effect of age and metal content of {\em old}
populations when compared to the optical -- is quite sensitive to the
presence of some amount of {\em young} stars, too (cfr. Burstein {\em et
al.} 1988). As a consequence, our newly obtained UV data provide a
direct tool to investigate the present and past role of star formation
processes in cluster galaxies. This information is of paramount
importance in the specific case of CL0939$+$4713, because -- for the
first time -- we can push our inquiry back in time up to an
intermediate-redshift cluster population of $\sim$5 billion years ago,
i.e. to some kind of ``missing ring'' in the current observational basis
for evolutionary studies (unlike clusters at z$\geq$2, the mid-UV region
of CL0939$+$4713 is not accessible from ground-based observations,
indeed).

For instance, the comparison with UV data of present-day galaxy clusters
provide a direct way of verifying whether signs of star formation
activity among member galaxies persist at the same/higher/lower level
when looking back at z$\sim$0.4 (at least as far as the two galaxy
samples are comparable). Moreover, when compared with suitable models of
populations (either purely old or hosting some recent star formation),
our data allow us to establish how recently UV-bright galaxies detected
by WFPC2 underwent their last episode of star formation. In the
following, analyses like those sketched above are applied to our F300W
photometry of CL0939$+$4713.

In particular, the availability of balloon-borne, mid-UV images of the nearby
Coma cluster obtained by Donas {\em et al.} (1995) make it obligatory to
(cautiously) compare these data with our HST magnitudes and colors of galaxies
belonging to CL0939$+$4713 (see Fig.~5).

In this respect, one should notice that the wavebands corresponding to
Donas {\em et al.}'s UV, optical magnitudes ($m_{UV}$, b) match closely,
when observing {\em local} galaxies, the wavebands imaged {\em at}
CL0939$+$4713 {\em restframe} by our own ($m_{300}$, m$_{702}$) HST magnitudes.
In other words, since we are imaging the two clusters at the same restframe
wavelengths, we are allowed to compare their populations taking into account
the relative distance modulus and a zeropoint shift (estimated to be 0.65~mag)
between HST and Donas {\em et al.}'s photometry alone (i.e. without applying
the usual K-correction). Unlike the preliminary, similar figure shown by Buson
{\em et al.} (1998), field galaxies are now made distinguishable from those
belonging to CL0939$+$4713, while objects lacking a redshift estimate have been
removed.

Actually, when comparing the UV properties of the two populations
superimposed in Fig.~5 and spaced in redshift by an amount
$\Delta$z$\sim$0.4, one should be fully aware of the heavy observational
biases and limitations involved.

First, the FOCA aperture used to image the nearby Coma cluster consists of a
circular aperture of about 1$^\circ$ in radius, while the three WF CCDs of the
WFPC2 camera cover a field of 75$''$$\times$75$''$ each. Even normalizing to
the same angular diameter distance, it turns out that the UV data of Coma come
from a FOV 18$\times$ larger, whose radius exceeds by many times the cluster
(X-ray) core radius (Briel {\em et al.} 1992); conversely, our HST images of
CL0939$+$4713 cover a region well within the cluster core radius
(r$_c\sim$3$'$; Schindler \& Wambsganss 1996). Secondly, despite a large
observational effort, the number of redshift measurements for alleged members
of CL0939$+$4713 is still confined to a few tens (Dressler {\em et al.} 1999),
thus implying that the number of UV-bright galaxies detected within the WFPC2
field (60 altogether) which can be assigned either to the cluster or the field
population by means of a reliable redshift estimate is necessarily tiny, as yet
(actually, as far as membership [{\em i.e.} redshift] determination is
concerned, UV-bright galaxies -- presumably showing emission lines in their
spectrum -- are favoured in comparison with random galaxies in the same field
and, as such, are the objects for which the present ambiguity could be more
easily removed by means of specific, follow-up observing programmes). Finally,
when seen at CL0939$+$4713 redshift, a significant fraction of a hypothetical
galaxy population with UV properties identical to those of Coma cluster
ellipticals would fall beyond the detection limit of the WFPC2 observations
presented here, being too red (see Fig.~5). This is confirmed by the very low
detection rate of early-type systems at z=0.4, as only one of 13 E/S0's within
the WFPC2 field listed as CL0939$+$4713 members by Dressler {\em et al.} (1999)
is detected in the UV.

As a consequence, the remarkably different portion of cluster population
sampled by the two kinds of UV experiments, together with the poor
absolute statistics of suitable cluster/field WFPC2 detections, prevent
us from drawing any significant conclusion about possible large-scale
population differences between CL0939$+$4713 and Coma in the UV. The
same is obviously true when comparing CL0939$+$4713 and its own field.
At this stage one can simply notice that c-m diagrams of Fig.~5 do not
offer any evidence in favor of the presence -- among the {\em
centrally-located} galaxy population of CL0939$+$4713 -- of objects much
bluer (i.e. UV-brighter) than the bluest star-forming galaxies pervading
the present-day Coma cluster. This, in turn, suggests that the excess of
star-forming objects noticed at z$\sim$0.4 does not imply an enhanced
star-formation rate in {\em individual} galaxies. However, we want to
stress once more that the possibility of quantifying the role of a
large-scale phenomenon such as the Butcher-Oemler effect in the mid-UV
is beyond the capabilities of the WFPC2 UV data of CL0939$+$4713
discussed here.

\section{COMPARISON WITH EVOLUTIONARY MODELS}

As already pointed out, one can complement the above analysis by
comparing the location within UV/optical c-m diagrams of our detected
galaxies with that of suitable population models. In this way one can
estimate how frequently individual cluster galaxies are affected by
major episodes of intervening star formation (up to continuous star
formation processes, obviously indistinguishable from a very recent
event). In the following we restrict our analysis to spheroidal galaxies
alone, i.e. avoiding young-population dominated spiral and/or irregular
galaxies. Owing to their normally low or nonexistent star formation,
ellipticals are indeed the best tracers of episodic star formation
events in clusters, such as might be caused by merging and/or tidal
disturbances, intergalactic gas stripping and ram pressure and
so on.

Such a comparison is shown in Fig.~6, where model spheroidal galaxies
representing passively evolving populations with/without the addition of some
amount of younger stars of different age are superimposed to our mid-UV/optical
c-m diagram for elliptical galaxies belonging to CL0939$+$4713, its own field,
and Coma (as seen at z=0.41), respectively. More precisely, models on the right
side of the figure represent aging single-burst populations at a fixed age of
10.8 Gyr (the estimated age of CL0939$+$4713 when adopting H$_0$=50 km s$^{-1}$
Mpc$^{-1}$, q$_0$=0 and z$_f$=5). Only three representative galaxy masses
(10$^{10}$~M$_\odot$, 10$^{11}$~M$_\odot$ and 10$^{12}$~M$_\odot$,
respectively) are shown here; a characteristic time $\tau$ corresponding to
0.1$\times$ the Hubble time has been adopted for the initial starburst.

The fact that the UV/optical color of uncontaminated model galaxies in Fig.~6
appears progressively redder and redder towards lower-mass systems should be
not surprising in view of the behavior of UV emission in old stellar
populations. Our adopted chemo-spectro-photometric models (see Tantalo {\em et
al.} 1996) indeed include the contribution of hot, evolved sources such as
Hot-HB and AGB-manqu\'e stars of high metallicity, {\em i.e.} the hot components
whose onset a late ages is responsible of the appearance of the well-known
UV-excess in ellitpical galaxies. Since the amplitude of this phenomenon
becomes higher and higher with growing galaxy mean metallicity (and total
mass), one can straightforwardly realize why old, passively evolved spheroidal
galaxies of higher total mass appear slightly bluer in their observed
($m_{300}-m_{702}$) color, an effect simply related to the higher level of
UV-excess contamination affecting the sampled spectral region (around 2100~\AA\
at z=0.41).

When recent episodes of star formation are added to the above models the
representative points of such ``contaminated'' populations move definitely
towards the left (blue) portion of the CMD. In particular, our computations
show the effect of adding to the model ellipticals recent `rectangular' bursts
with slightly different efficiencies $\nu_b$ (within 1-3\%) and constant
duration (10$^8$ yr), centered at a time $\tau_b$ of 0.1, 0.4 and 0.8 Gyr
before the epoch recorded by our observations respectively.

One can recognize three major features when comparing UV observations and
models: \begin{enumerate} \item Purely passively evolved populations fall
beyond the detection capabilities of the UV observations discussed here (both
WFPC2 and FOCA). The same is true for intermediate/low luminosity ellipticals
redder than $m_{300}-m_{702}\sim$3 and $m_{300}-m_{702}\sim$2 respectively, an
observational bias giving rise to a growing {\em spurious} gap between less
massive galaxy models and detected objects. \item Secondary bursts experienced
by galaxies earlier than 0.8 Gyr before the observing epoch are fully
reabsorbed in terms of restframe mid-UV/optical colors, independently of galaxy
mass and luminosity; this implies that {\em all spheroidal galaxies UV-bright
enough to be detected in our HST F300W (as well as balloon-borne) images did
host a starburst in the near past (typically 100-300~Myr earlier) or,
alternatively, undergo a low-level, continuous star formation activity.} \item
Among the numerous UV-bright, early-type galaxies seen in Coma (and not at the
center of CL0939$+$4713), signs of very recent (and thus presumably frequent)
star formation activity is shown preferentially by low-luminosity objects. In
this respect, it is worth mentioning that the existence of this kind of
continuous rejuvenation for Coma cluster low-luminosity early-type galaxies
(i.e. a series of overlapping short bursts) is largely supported also by recent
spectroscopic observations (Caldwell \& Rose 1998) showing that a significant
fraction of these faint members do host a young population superposed to older
stars. \end{enumerate}

\section{LOOKING AT CL0939$+$4713 IN THE FAR-UV}

As already pointed out, our HST observations aimed also at exploring the
far-UV portion of the {\em restframe} energy distribution of galaxies at
z$\sim$0.4. Actually, the presence of a far-UV excess (upturn) in
metal-rich, evolved populations is known since the early epochs of UV
space astronomy (Code 1969; Code \& Welch 1979; Bertola {\em et al.}
1980, 1982, 1986; Oke {\em et al.} 1981) and {\em the goal of recording
its onset and subsequent evolution at intermediate redshift is still a
major observational challenge} (see Greggio \& Renzini 1999 and
O'Connell 1999 for comprehensive, recent reviews).

In particular, on the basis of current galaxy evolutionary models (e.g.
Bressan {\em et al.} 1994; Tantalo {\em et al.} 1996), one should expect
that -- owing to the vanishing of post-HB evolutionary paths (Post
Early-AGB stars, AGB-Manqu\'e stars) which generate most of the
present-day UV-bright stars -- the UV properties of old populations in
giant ellipticals do show a dramatic change, when observed to a proper
lookback time (which could well fall around CL0939$+$4713 redshift). In
other words, looking back in time a few billion years, one should
witness some kind of `switching off' of the UV-upturn seen in
present-day giant ellipticals, thus recording restframe 1550$-$V colors
much redder than their present value.

Unfortunately, the restframe far-UV emission (F218W) of almost the
totality of galaxies detected in the restframe mid-UV (F300W) could not
be recorded and even the upper limits to the level of the far-UV flux in
individual ellipticals at z=0.4 implied by our WFPC2 observations are
too high to provide the sought astrophysical constraints. More
precisely, the upper limit to the UV emission imposed by adding F218W
counts in the {\em total} frame area occupied by the {\em whole sample}
of optically-bright ellipticals in the corresponding F702W frames,
implies that one cannot detect objects (at 3~$\sigma$ limit) whose
restframe UV/optical color is higher (i.e. {\em redder}) than
(1550--V)=1.0. Such a (blue) color can be reached among present-day
galaxies only by actively star-forming systems, while even the nearby
UV-brightest old populations in ellipticals are not bluer than
(1550--V)$\sim$2 and, as such, could not be detected ({\em cfr.}
Burstein {\em et al.} 1988).

As a consequence, the set of (F218W) UV data discussed here are
inadequate to establish the existence and amplitude of the UV-upturn
phenomenon typical of present-day evolved populations at CL0939$+$4713
lookback time. This outcome is consistent with similar unconclusive
results reached by other groups who observed clusters at similar
distance with the same instrumental configuration (e.g. Renzini 1996).
This conclusion -- though negative -- does not imply that such a kind of
investigation is {\em strictly} beyond HST capabilities, however, and is
presently pushing other groups to explore alternative observing
approaches. In this respect, the very recent positive result of Brown
{\em et al.} (1998) -- who were able to sample the UV-upturn of {\em a
few} ellipticals belonging to a cluster of similar redshift (Abell~370;
z=0.375) by combining two FOC long-pass filters -- holds the hope of
exploiting successfully HST in the near future, in particularly with the
planned installation of the Advanced Camera for Surveys (ACS).

\section{CONCLUSIONS}

This work can be considered as the first ``journey'' into the restframe
UV of an {\em intermediate-redshift}, rich galaxy cluster. Making use of
HST, we were able to detect in a single WFPC2 field tens of UV-bright
sources belonging either to CL0939$+$4713 or to its close
foreground/background. Newly obtained UV data have been combined with
pre-existing HST and ground-based optical images to derive unprecedented
UV/Optical c-m and two-color diagrams. While irregular and (luminous)
early-type galaxies are well-confined to specific color ranges within
such diagrams, spirals do exhibit quite a large variability in color.

In order to explore possible evolutionary effects, HST data of
CL0939$+$4713 have been finally compared with balloon-borne UV data of
the {\em local} Coma cluster. Although an {\em exhaustive} comparison of
the hottest populations of nearby and intermediate-redshift clusters has
to wait until deeper, wide-field UV surveys of distant clusters will
become available, the limited datasets discussed here do provide
interesting pieces of evolutionary information. For instance no hints of
a fast evolution in the UV from z$\sim$0.4 to present time (in the sense
of cluster galaxies at z$\sim$0.4 hosting a star-formation activity
dramatically higher than their present-day counterparts) are found.

\clearpage


\begin{deluxetable}{ccccccc}
\tablecaption{Abell851 F300W Photometry}
\tablewidth{12cm}
\tablehead{
\colhead{ID} & \colhead{S97 Number} & \colhead{X} & \colhead{Y}
& \colhead{$m_{300}$} & \colhead{$m_{300}-m_{702}$}& \colhead{T-Type}\\
\colhead{(1)} & \colhead{(2)} & \colhead{(3)} & \colhead{(4)} &
\colhead{(5)} & \colhead{(6)} & \colhead{(7)}
}
\startdata
  1 &  125 & 1256 &  326 & 21.53 & -0.81 & -99\\
  2 &  660 & 1427 & 1005 & 23.76 & -0.99 & -99\\
  3 &  301 &  221 &  538 & 23.33 & -1.15 & -99\\
  4 &  706 &  940 & 1118 & 23.59 & -2.01 & -99\\
  5 &  892 &  856 & 1505 & 23.63 & -1.18 & -99\\
  6 &  768 & 1356 & 1265 & 24.11 & -1.78 & -99\\
  7 &  805 & 1418 & 1356 & 23.78 & -1.77 & -99\\
  8 &  457 &  156 &  740 & 24.03 & -1.40 & -99\\
  9 &   64 & 1225 &  256 & 24.61 & -0.38 & -99\\
 10 &  873 & 1143 & 1463 & 23.85 & -0.71 & -99\\
 11 &   72 &  969 &  271 & 23.34 & -1.39 & -99\\
 12 &  866 & 1159 & 1453 & 24.02 & -0.91 & -99\\
 13 &  674 & 1155 & 1027 & 22.97 & -1.35 & -99\\
 14 &  136 & 1238 &  341 & 23.78 & -0.95 & -99\\
 15 &  178 & 1025 &  405 & 23.23 & -0.21 & -99\\
 16 &  760 & 1048 & 1248 & 24.07 & -0.66 & -99\\
 17 &  399 &  285 &  635 & 23.92 &  3.31 &  -5\\
 18 &  752 & 1001 & 1230 & 22.60 & -1.02 &  -4\\
 19 &  455 & 1392 &  727 & 23.22 &  0.46 &  -3\\
 20 &  425 & 1046 &  678 & 23.80 &  2.89 &  -2\\
 21 &  273 & 1314 &  489 & 23.27 &  3.05 &  -2\\
 22 &   35 &  871 &  238 & 22.60 & -0.60 &  -2\\
 23 &  738 & 1444 & 1211 & 22.92 & -0.40 &   1\\
 24 &  427 &  152 &  670 & 23.88 &  3.37 &   1\\
 25 & 3017 &   44 &  411 & 23.53 &  0.01 &   1\\
 26 &  456 &  117 &  736 & 23.51 & -0.07 &   2\\
 27 &  735 &  984 & 1165 & 23.58 & -0.26 &   3\\
 28 &   43 &  667 &  242 & 23.37 &  0.81 &   4\\
 29 &  215 &  634 &  430 & 23.29 &  1.26 &   4\\
 30 &  369 &  830 &  605 & 21.79 & -0.34 &   4\\
 31 &  763 & 1425 & 1245 & 23.09 &  0.19 &   4\\
 32 &  497 &  139 &  768 & 20.66 & -1.06 &   4\\
 33 &  258 &  845 &  484 & 21.52 &  0.05 &   4\\
 34 & 3006 &   77 &  389 & 23.29 &  0.15 &   5\\
 35 &  824 &  940 & 1376 & 23.58 &  0.51 &   5\\
 36 &  422 & 1341 &  681 & 21.91 & -0.30 &   5\\
 37 &  788 & 1420 & 1315 & 23.16 & -0.79 &   5\\
 38 &   65 & 1218 &  271 & 23.15 & -0.65 &   5\\
 39 &  321 & 1089 &  553 & 23.30 & -0.82 &   6\\
 40 &  224 &   56 &  367 & 23.03 &  1.96 &   6\\
 41 &  601 & 1287 &  905 & 21.65 & -1.23 &   6\\
 42 &  261 &  331 &  495 & 22.75 & -0.24 &   6\\
 43 &  540 & 1150 &  812 & 23.28 &  0.82 &   6\\
 44 &   84 &  928 &  285 & 23.24 & -0.58 &   6\\
 45 &  123 &  318 &  293 & 23.87 &  1.17 &   6\\
 46 &  232 &  760 &  462 & 23.07 & -0.68 &   6\\
 47 &  806 &  904 & 1337 & 23.84 & -0.48 &   7\\
 48 &  742 & 1286 & 1208 & 23.86 &  1.24 &   7\\
 49 &  809 & 1048 & 1348 & 23.18 & -0.74 &   7\\
 50 &  736 & 1028 & 1199 & 22.91 & -0.29 &   7\\
 51 &  339 &  937 &  579 & 23.05 & -0.47 &   7\\
 52 &  489 &  767 &  766 & 22.61 &  0.33 &   7\\
 53 &  851 & 1117 & 1427 & 23.19 & -0.74 &   7\\
 54 &  122 &  246 &  308 & 21.86 &  0.30 &   7\\
 55 &   36 &  300 &  229 & 21.69 &  0.26 &   7\\
 56 &  342 &  666 &  588 & 23.49 & -0.82 &  10\\
 57 &  750 & 1125 & 1229 & 22.91 & -1.29 &  10\\
 58 &  829 & 1122 & 1394 & 23.40 & -0.60 &  10\\
 59 &  191 &  133 &  422 & 23.29 & -0.43 &  10\\
 60 &  230 &  734 &  458 & 20.44 & -1.18 &  10\\
\enddata
\tablecomments{Col.~(1): Object identifier; Col.~(2): Cross identification
of objects against the catalog in Smail et al. (1997); Col.~(3): X centroid
on the mosaicked F300W frame in pixels; Col.~(4): Y centroid on the
mosaicked F300W frame in pixels; Col.~(5): 1\farcs2 diameter aperture
magnitude (STMAG) measured in the F300W waveband; Col.~(6): Color in a 1
\farcs2 diameter aperture ($m_{702}$ magnitude being measured in the F702W
waveband); Col.~(7): Standard T-Type; an undefined entry is given as -99.}
\end{deluxetable}

\clearpage

\begin{deluxetable}{cccccccc}
\tablecaption{Abell851 F218W Photometry}
\tablewidth{15cm}
\tablehead{
\colhead{ID} & \colhead{S97 Number} & \colhead{X} & \colhead{Y} & \colhead{
$m_{218}$} & \colhead{$m_{218}-m_{300}$} & \colhead{$m_{218}-m_{702}$} &
\colhead{T-Type}\\
\colhead{(1)} & \colhead{(2)} & \colhead{(3)} & \colhead{(4)} &
\colhead{(5)} & \colhead{(6)} & \colhead{(7)} & \colhead{(8)}
}
\startdata
 32 &  497 &  139 &  768 & 20.66 & -0.56 & -1.62 & 4\\
 60 &  230 &  734 &  458 & 20.44 & -0.45 & -1.63 & 10\\
\enddata
\tablecomments{Col.~(1): Object identifier; Col.~(2): Cross identification
of objects against the catalog in Smail et al. (1997); Col.~(3): X centroid
on the mosaicked F300W frame in pixels; Col.~(4): Y centroid on the
mosaicked F300W frame in pixels; Col.~(5): 1\farcs2 diameter aperture
magnitude (STMAG) measured in the F218W waveband; Col.~(6): Color in a 1
\farcs2 diameter aperture ($m_{300}$ magnitude being measured in the F300W
waveband); Col.~(7): Color in a 1\farcs2 diameter aperture ($m_{702}$
magnitude being measured in the F702W waveband); Col.~(8): Standard
T-Type.}
\end{deluxetable}



\clearpage
\plotone{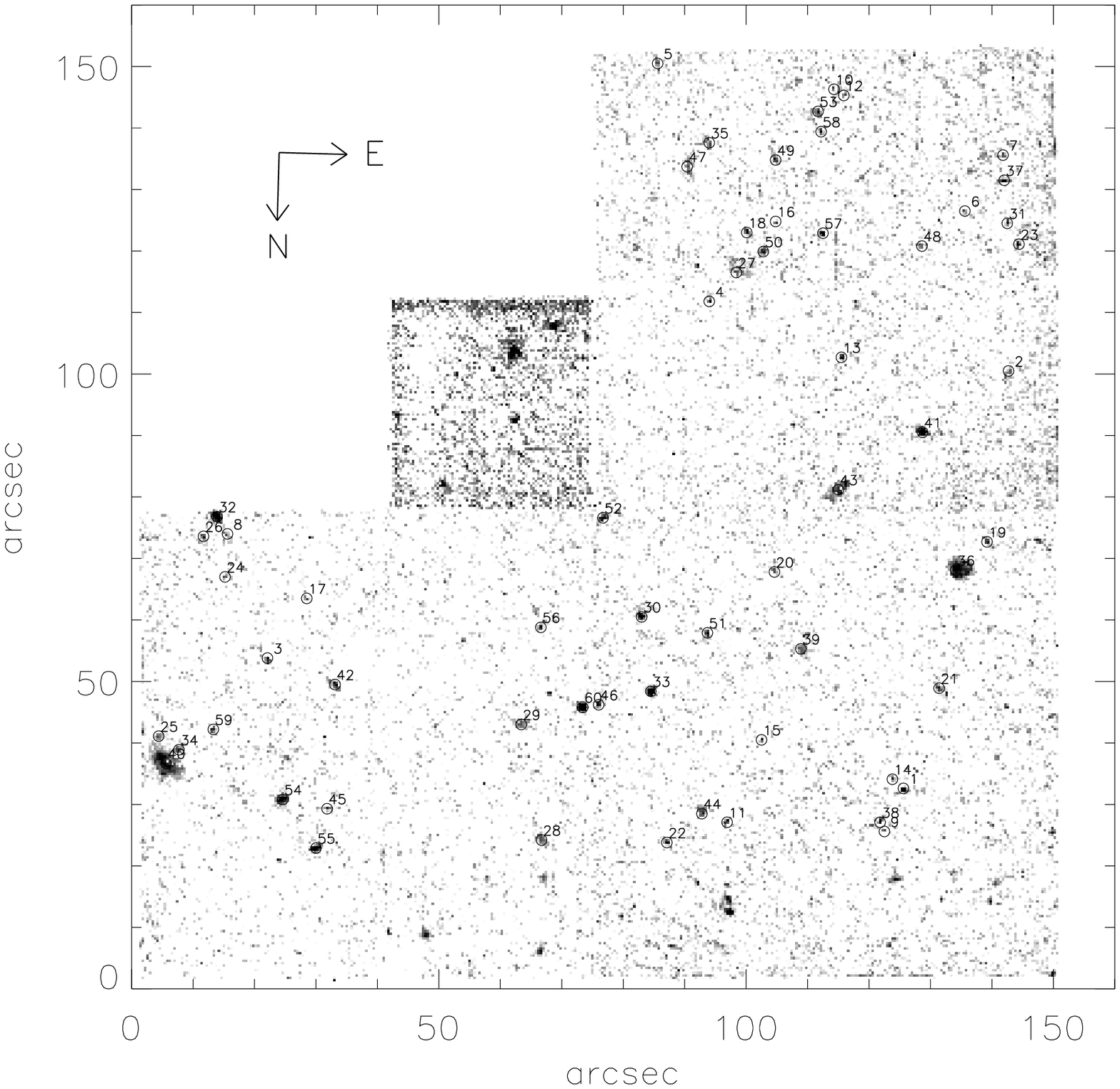}
\figcaption{Five-orbit, co-added near-UV WFPC2 frame of CL0939$+$4713
(imaged through the F300W filter). A logarithmic display scale has been
adopted. Detected galaxies are marked with a circle and the sequence
number adopted in Table~1 and 2. Note that galaxies identified as source
no.~32 and no.~60, respectively are the only objects detected also in
mid-UV (F218W) frames. The PC1 field has been excluded from our
analysis, due to its higher noise.}

\clearpage
\plotone{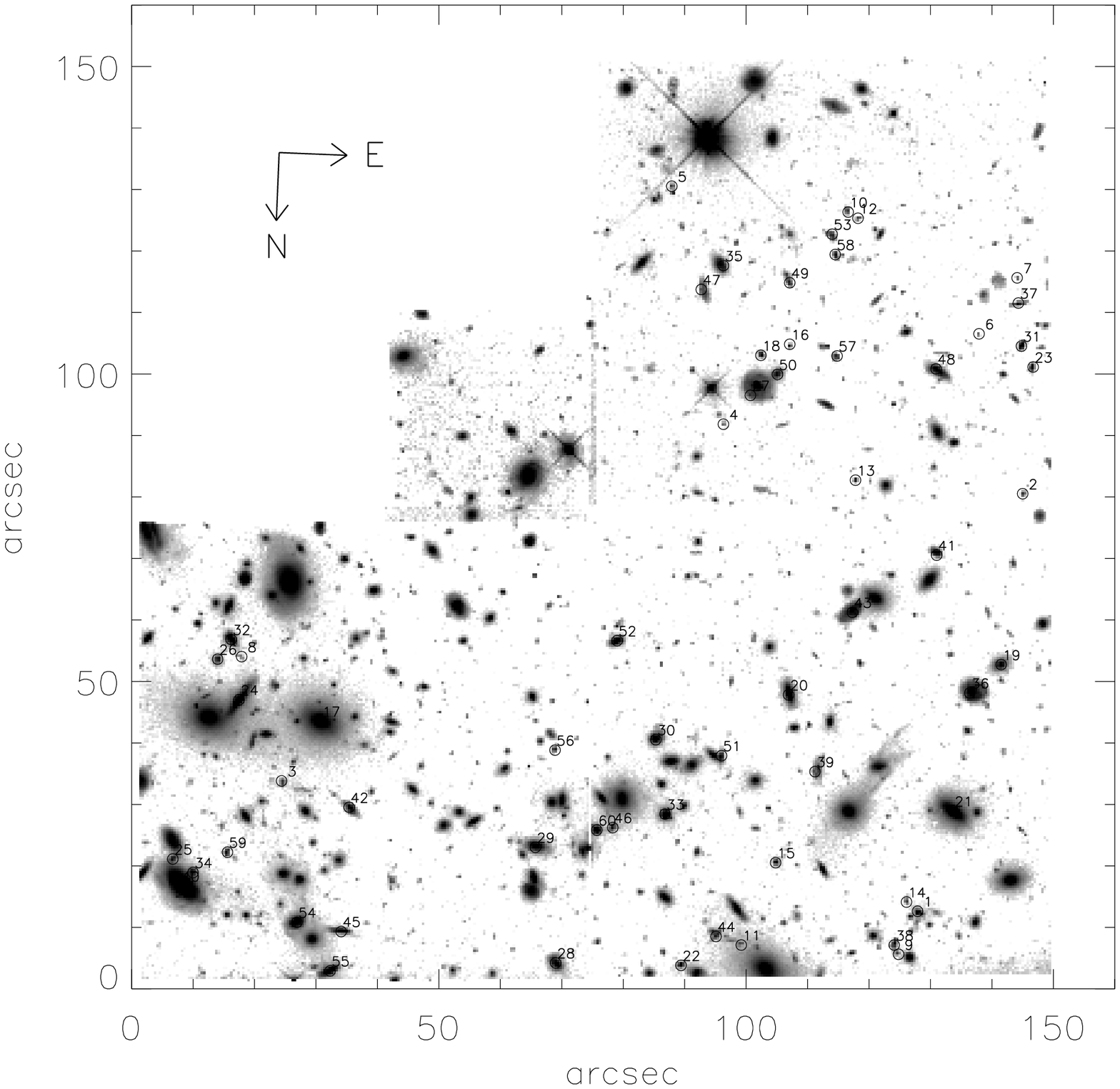}
\figcaption{Comparison optical (F702W) WFPC2 frame of CL0939$+$4713.
Again, a logarithmic display scale has been adopted. Sources detected in
the F300W waveband are marked as in previous figure for
cross-identifying purposes. The fields of optical and UV images do not
match perfectly, being shifted of $\sim$20~arcsec approximately along
the North-South direction.}

\clearpage
\plotone{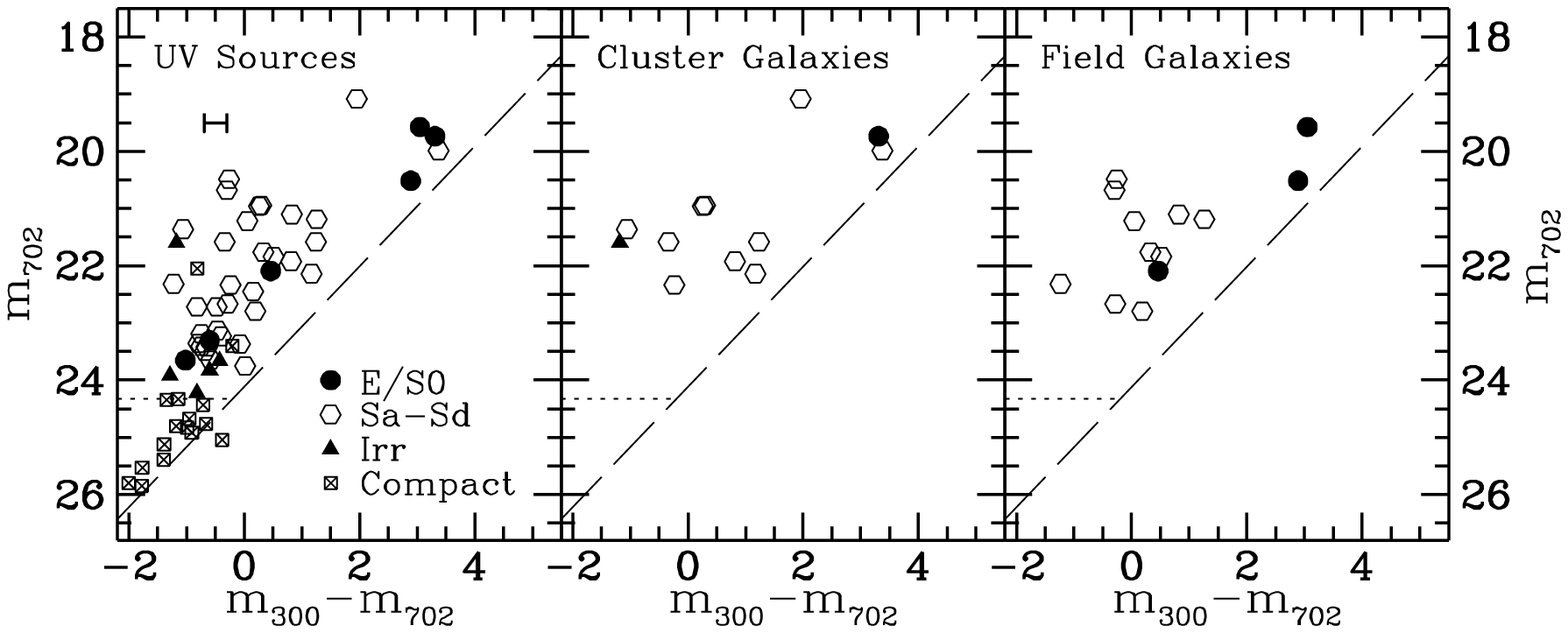}
\figcaption{{\em Left Panel:} Overall c-m ($m_{702}$ {\em vs.}
[$m_{300}-m_{702}$]) diagram of the sources detected in the WFPC2 frames
centered on CL0939$+$4713. The ($m_{300}-m_{702}$) color derived from our
aperture photometry (0.6$''$ radius) is compared with {\em integrated}
$m_{702}$ magnitudes given by Smail {\em et al.} (1997) after applying a
zeropoint correction of 0.85 mag to take correctly into account the value of
the detector gain and of the adopted aperture (namely,
$\Delta$mag~=~0.85~=~2.5~log[2]$+$0.1 where, with reference to the paper of
Holtzmann et al. [1995], the two terms compensate for a 2$\times$ lower gain
value (g=7) and infinite aperture, respectively). A key to the morphological
types is inserted. Shaded line and dotted line represent the detection limit
imposed by our UV images and the limit magnitude adopted by Smail {\em et al.}
for their (optical) morphological classification, respectively. Formal errors
of our HST UV photometry are shown by the representative error bar. Formal
errors for HST optical measurements are negligible. {\em Central Panel:} the
same for spectroscopically confirmed members of CL0939$+$4713 alone. {\em Right
Panel:} The same for confirmed field (either foreground or background) galaxies
projected onto the CL0939$+$4713 WFPC2 field of view.}

\clearpage
\epsscale{0.7}
\plotone{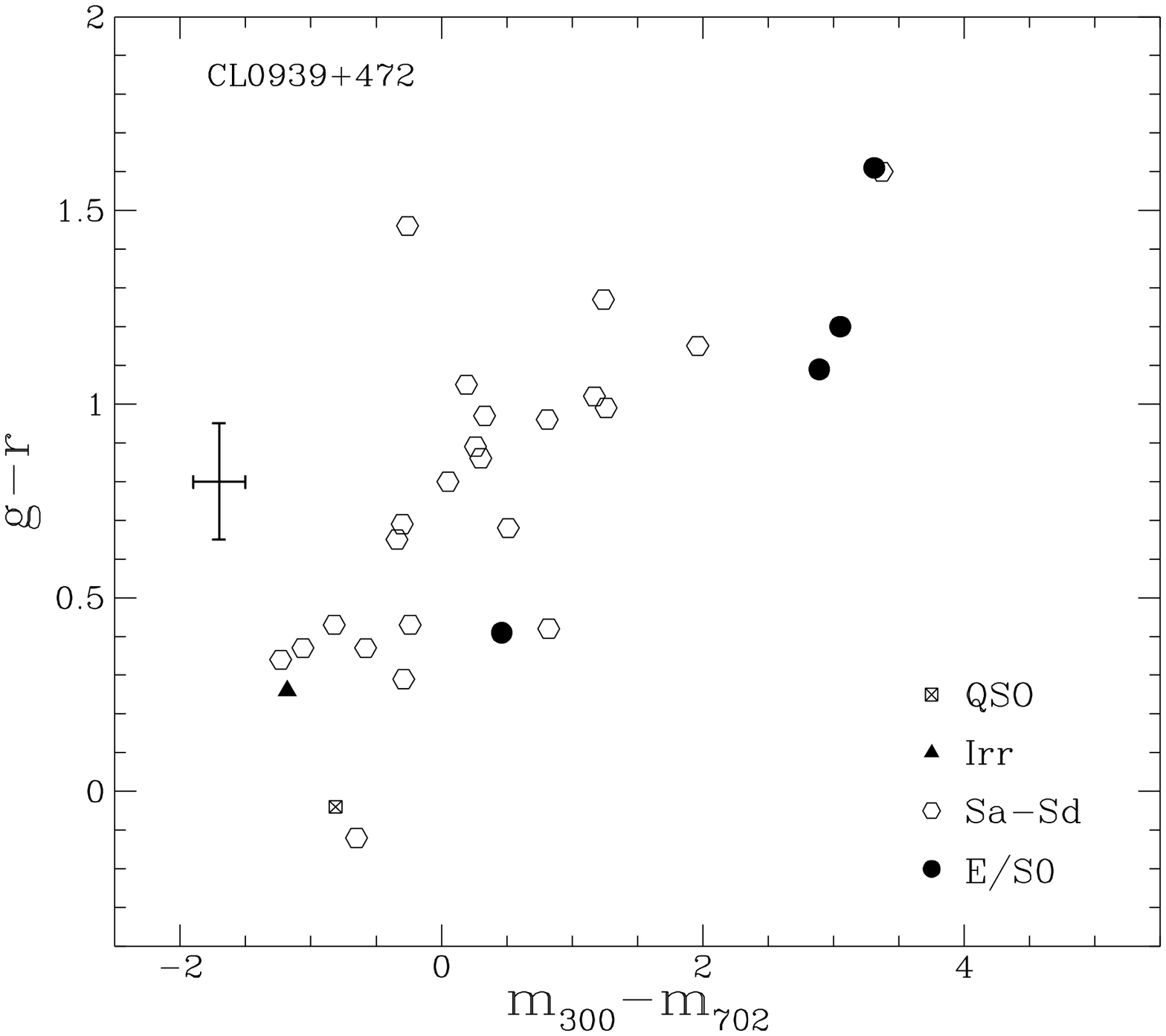}
\figcaption{Two-color, ([$g-r$] {\em vs.} [$m_{300}-m_{702}$]) diagram
of all sources simultaneously detected in the UV within the
CL0939$+$4713 WFPC2 field and photometered from ground by Dressler \&
Gunn (1992). The ($m_{300}-m_{702}$) color always refers to an aperture
of 0.6$''$ in radius. A key to the morphological types is inserted.
Error bars combine formal errors of our HST UV photometry and expected
systematic errors of optical, ground-based photometry.}

\clearpage
\epsscale{1}
\plotone{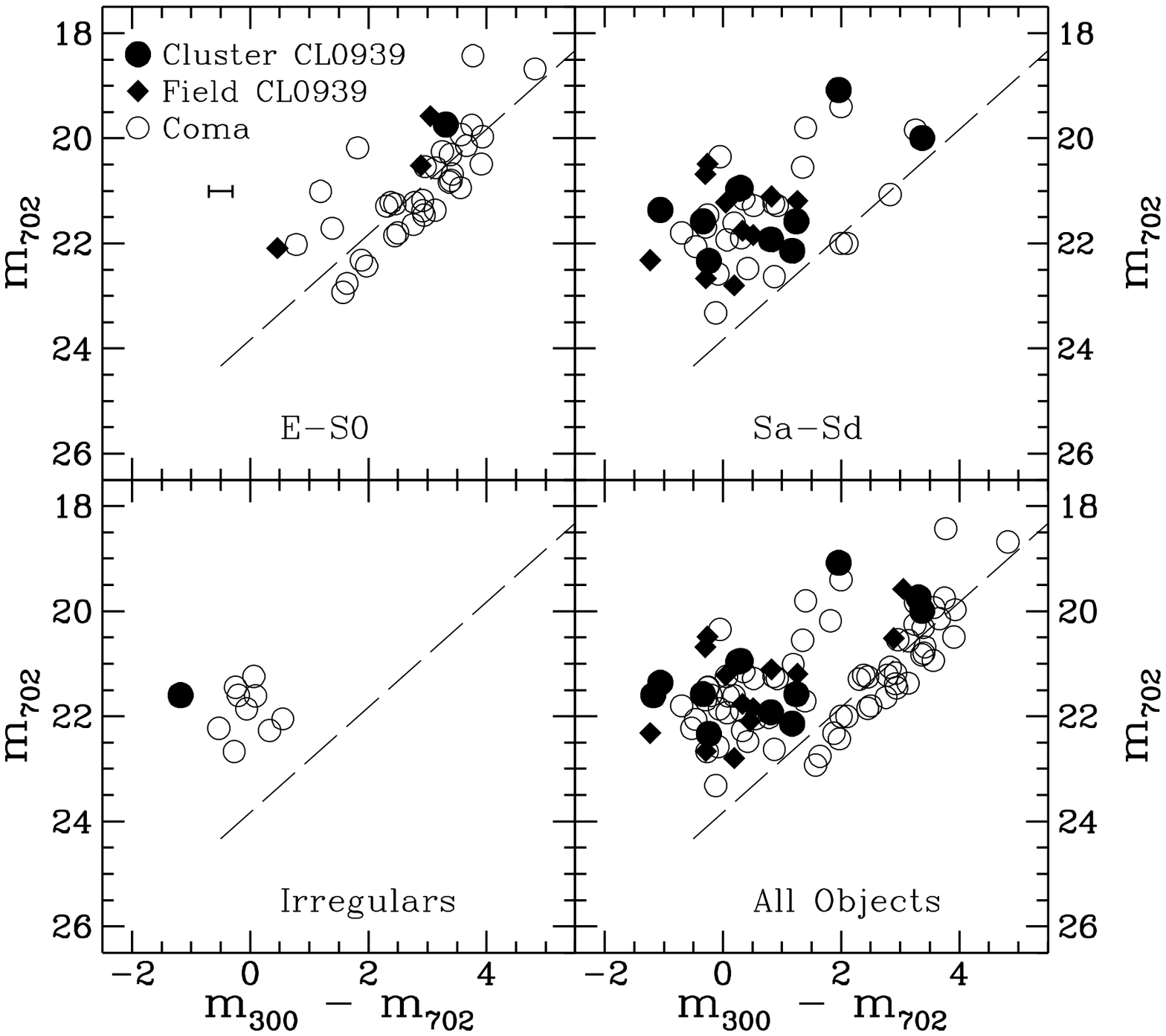}
\figcaption{UV/optical c-m diagrams superimposing CL0939$+$4713 member galaxies
(filled circles), CL0939$+$4713 field galaxies (filled diamonds) and Coma
cluster galaxies (open circles), splitted into morphological classes. A
relative distance modulus $\Delta\mu$=6.47 is assumed to scale down UV/optical
magnitudes (m$_{UV}$, b) of Donas {\em et al.} for Coma to our observed HST
magnitudes (m$_{300}$, m$_{702}$). A further amount of 0.65 mag has been
subtracted from b magnitudes to remove the zeropoint mismatch between the two
photometric systems (thus implying m$_{702}$~=~[b$-$0.65]$+$6.47, as well as
[$m_{300}-m_{702}$]~=~[m$_{UV}-$b$]+$0.65). Only UV-bright galaxies for which
an optical morphological classification is available are included. The shaded
line represents the detection limit of WFPC2 observations. Formal errors of our
HST UV photometry are shown by the representative error bar. Formal errors for
HST optical measurements are negligible.}

\clearpage
\epsscale{0.7}
\plotone{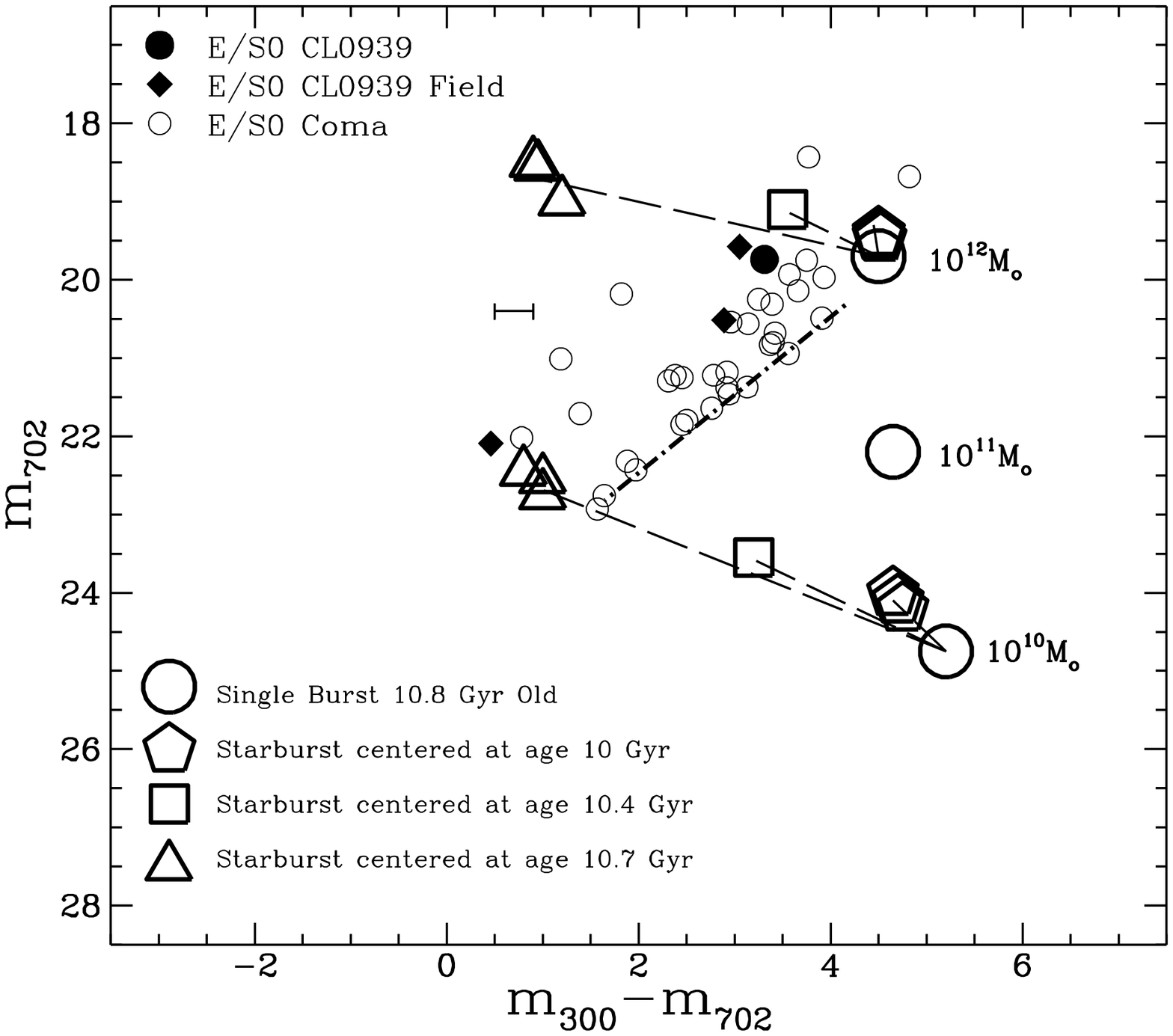}
\figcaption{Models showing the expected location of both passively evolved and
young-population contaminated ellipticals, superimposed to the UV/ optical c-m
diagram for early-type galaxies of both CL0939$+$4713 and Coma clusters. Both
models and galaxy magnitudes have been scaled to z=0.41. The meaning of
symbols, both for data and models, is given by the key in figure. On the right
side are represented galaxies at a fixed age of 10.8 Gyr with mass of
10$^{10}$~M$_\odot$, 10$^{11}$~M$_\odot$ and 10$^{12}$~M$_\odot$, respectively,
which consist of an aging, uncontaminated single-burst population. Dashed lines
connect both the highest and lowest mass models to their expected location in
case they have experienced recent `rectangular' bursts having constant duration
(10$^8$ yr) and slightly different efficiencies $\nu_b$ (within 1-3\%),
centered at a time $\tau_b$ of 0.1, 0.4 and 0.8 Gyr before the epoch recorded
by our observations, respectively. Multiple symbols correspond to the above
slightly different efficiencies of the recent burst. The dotted-dashed line
represents the FOCA detection limit for UV observations of Coma cluster
galaxies. Formal errors of our HST UV photometry are shown by the
representative error bar. Formal errors for HST optical measurements are
negligible.}

\end{document}